\documentclass[prb,twocolumn,showpacs]{revtex4}
\usepackage{graphicx}

\begin{document}

\title{
   Landau level mixing in the $\nu=5/2$ fractional quantum Hall state}

\author{
   Arkadiusz W\'ojs,$^{1,2}$ and
   John J. Quinn$^2$}

\affiliation{
   \mbox{
   $^1$Institute of Physics, Wroc{\l}aw University of Technology,
       Wybrze\.ze Wyspia\'nskiego 27, 50-370 Wroc{\l}aw, Poland}\\
   $^2$Department of Physics, University of Tennessee, 
       Knoxville, TN 37996, USA}

\begin{abstract}
The $\nu=5/2$ fractional quantum Hall state is studied numerically, 
directly including the effects of electron scattering between 
neighboring Landau levels.
Significant reduction of the excitation gap caused by the LL mixing
explains the discrepancy between earlier calculations and experiments.
On the other hand, LL mixing also considerably reduces overlaps 
with the Moore--Read wavefunction, raising a question of the actual 
realization of nonabelian quasiparticles in present experiments.
\end{abstract}
\maketitle

\section{Introduction}

Discovery\cite{Willett87} of the fractional quantum Hall (FQH)
effect in a half-filled first excited Landau level (LL$_1$) 
demonstrated the possibility for incompressible quantum 
liquid (IQL) states outside of the ``standard'' hierarchy
\cite{Laughlin83,Haldane83} described below.
Laughlin states\cite{Laughlin83} occur at the fractional LL 
fillings $\nu\equiv2\pi\varrho\lambda^2=(2p+1)^{-1}$ (where 
$\varrho$ is electron concentration, $\lambda$ is the magnetic 
length, and $p$ is an integer).
The simple form of Laughlin many-body wavefunctions suggested 
that their low energy could be understood in terms of the 
avoidance of pair states with the smallest relative pair 
angular momentum (and largest repulsion).\cite{Haldane87}
The quasiparticles (QPs) of the Laughlin states can (under 
some conditions\cite{hierarchy}) form ``daughter'' IQL states 
with their own QPs, giving rise to an entire IQL hierarchy.
\cite{Haldane83}
The most stable IQL states, occurring at $\nu=s(2ps\pm1)^{-1}$ 
($s$ being another integer), appear naturally in Jain's composite 
fermion (CF) model\cite{Jain89} involving the concepts of 
flux attachment and an effective magnetic field.
All Laughlin and Jain states are characterized by odd-denominator 
filling fractions $\nu$ and fractional QP charge 
$q=e(2ps\pm1)^{-1}$.

For the even-denominator IQL state observed at $\nu={5\over2}$, 
Moore and Read (MR) proposed\cite{Moore91} a different, paired 
wavefunction, and predicted that its QPs obeyed nonabelian 
statistics.
The MR state has been studied in great detail\cite{Wen93} and 
interpreted by two complementary\cite{Read00} pictures: as a 
Laughlin state of tightly bound electron pairs\cite{Greiter91,3body} 
or a superfluid of weakly bound CF pairs.\cite{Jain00,Toke06}

The first numerical calculations for interacting electrons in 
a partially filled LL$_1$ were carried out by Morf.\cite{Morf98}
They seemed to confirm that a half-filled LL$_1$ has a 
spin-polarized incompressible ground state accurately described 
by the MR wavefunction.
However, the subsequent experiments\cite{Pan99,Pan01,%
Eisenstein02,Xia04} revealed minute excitation gaps 
$\Delta\sim0.1-0.45$~K in the real $\nu={5\over2}$ states, 
up to $20\times$ smaller than predicted from numerics.
The fact that only a small part of this discrepancy could 
be attributed to the finite width of the quasi-2D electron 
layer or to the weak disorder posed a challenge for the
theoretical models.
It also raised a fundamental question of whether the actual, 
experimentally realized $\nu={5\over2}$ states are indeed
adequately described by the MR wavefunction.
It has become very important in the context of topological 
quantum computation, whose recent proposals
\cite{Kitaev03,DasSarma05,Bonesteel05} take advantage of the 
nonabelian QP statistics.
This fundamental question is the main subject of our present 
paper.

We directly include LL mixing in exact diagonalization by adding 
states containing a single cyclotron excitation to the Hilbert 
space of a partially filled spin-polarized LL.
This can be regarded as a first-order approximation in the 
Coulomb-to-cyclotron energy ratio $\beta$.
We apply this procedure to the $\nu={1\over3}$, ${2\over5}$, and 
${5\over2}$ states and evaluate the excitation energy gaps $\Delta$ 
and (in the last case) the overlap $\xi$ with the MR wavefunction.
At $\beta\sim1$ (relevant for the experiments at $\nu={5\over2}$), 
where our first-order results are only qualitative, we find a 
strong reduction in both $\Delta$ and $\xi$.
Unfortunately, the calculations including excitations to higher 
order in $\beta$ are beyond our capabilities.
Our conclusion concerning the gap reduction gives qualitative 
support to the work of Morf and d'Ambrumenil,\cite{Morf03} who 
however included LL mixing in a different way,\cite{Aleiner95} 
using a screened interaction that is strictly valid only for 
$\nu\gg1$.
Furthermore, our prediction of strong deviation from the MR 
wavefunction supports the recent proposals\cite{Stern06} for 
new experiments aimed at determining directly the QP statistics 
at $\nu={5\over2}$ (rather than merely at determining the 
incompressibility).

\section{Model}

We use Haldane's spherical geometry,\cite{Haldane83} convenient 
for the exact study of liquid states with short-range correlations.
On a sphere of radius $R$, the normal magnetic field $B$ is 
produced by a Dirac monopole of strength $2Q=4\pi R^2 B/\phi_0$, 
defined here in the units of flux quantum $\phi_0=hc/e$.
Using a magnetic length $\lambda=\sqrt{\hbar c/eB}$, this can be 
rewritten as $Q\lambda^2=R^2$.
The series of LLs labeled by $n=0$, 1, 2, \dots\ are represented 
by shells of angular momentum $l=Q+n$ and degeneracy $g=2l+1$.
The cyclotron energy is $n\hbar\omega_c=\hbar eB/\mu c$ (counted 
from the lowest LL), where $\mu$ is the effective mass.
The orbitals $\psi_{nm}(\theta,\phi)$ are called monopole 
harmonics.

The $\mathcal{N}$-electron Hamiltonian matrix is calculated 
in the configuration-interaction basis 
$\left|i_1,\dots,i_\mathcal{N}\right>$. 
Here, the composite indices $i=[n,m,\sigma]$ also include spin,
and the expressions for two-body Coulomb matrix elements 
(also in layers of finite widths $w$) can be found for example 
in Ref.~\onlinecite{trions}.
At a given LL filling (defined by $\mathcal{N}$ and $g$, the 
basis states can be classified by $\Delta\mathcal{E}=\hbar
\omega_c\sum_k n_k-\mathcal{E}_{\rm min}$, i.e., the total 
cyclotron energy measured from the lowest possible value 
$\mathcal{E}_{\rm min}$ allowed by the Pauli exclusion principle
(e.g., $\mathcal{E}_{\rm min}=0$ at $\nu\le2$, or 
$\mathcal{E}_{\rm min}=(\mathcal{N}-2g)\hbar\omega_c$ at 
$2<\nu\le4$).
Alternatively, the total number of cyclotron excitations 
$K=\Delta\mathcal{E}/(\hbar\omega_c)$ can be defined.

\begin{figure}
\includegraphics[width=3.4in]{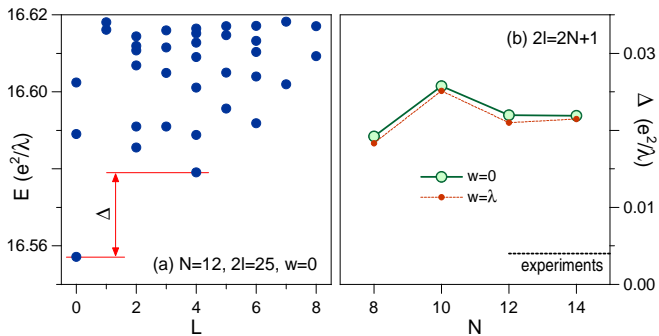}
\caption{(color online)
   (a) Energy spectrum (energy $E$ as a function of angular
   momentum $L$) of $N=12$ electrons in a LL shell of 
   angular momentum $l=25/2$, corresponding to an isolated, 
   half-filled first excited LL.
   (b) Excitation gap $\Delta$ extracted from the spectra
   similar to (a), plotted as a function of the electron 
   number $N$;
   $w$ is the width of the 2D electron layer and
   $\lambda$ is the magnetic length.}
\label{fig1}
\end{figure}

The exact numerical diagonalization in the Hilbert space 
restricted to $K=0$ means including Coulomb scattering within 
only one, partially filled LL, and neglecting the LL mixing.
For $\nu={5\over2}$ this reduces the $\mathcal{N}$-particle 
problem to $N=\mathcal{N}-2g$ electrons confined to an isolated 
LL$_1$.
A typical numerical spectrum is shown in Fig.~\ref{fig1}(a). 
The non-degenerate ground states with a gap generally appear 
in finite-size systems with even values of $N$ (they are 
known to be paired\cite{Moore91}) at $2l=2N+1$ or 
(equivalent via the $N\rightarrow g-N$ particle-hole conjugation) 
at $2l=2N-3$, both extrapolating to $N/g\rightarrow{1\over2}$
for large $N$.
As shown in Fig.~\ref{fig1}(b), the excitation gap $\Delta$ 
rather weakly depends on $N$, allowing one to estimate the 
value $\Delta\approx0.02\,e^2/\lambda$ for an infinite (planar) 
system.\cite{Morf98,Toke06}

Unfortunately, this value is not confirmed by the experiments.
The gaps measured from the thermal activation of longitudinal
conductance range from 0.001 to $0.004\,e^2/\lambda$, depending 
on the electron mobility,\cite{Pan99,Pan01,Eisenstein02,Xia04} 
with extrapolation to a disorder-free system not exceeding 
$\sim0.006\,e^2/\lambda$.
As shown in Fig.~\ref{fig1}(b) for $w=\lambda$ (i.e., $w=11.4$~nm 
at $B=5$~T), this discrepancy cannot be explained by a finite 
width of the electron layer.

An obvious advantage of the $K=0$ approximation is that 
calculations can be done for sufficiently large values 
of $\mathcal{N}$ to eliminate finite-size errors.
It could be trivially justified by a small ratio of Coulomb 
and cyclotron energies, 
$\beta=(e^2\lambda^{-1})/(\hbar\omega_c)=\lambda/a_{\rm B}$ 
(with the Bohr radius $a_{\rm B}=\hbar^2/\mu e^2$).
However, $\beta>1$ at the fields $B\sim5$~T typically used 
in FQH experiments at $\nu={5\over2}$.

\begin{figure}
\includegraphics[width=3.4in]{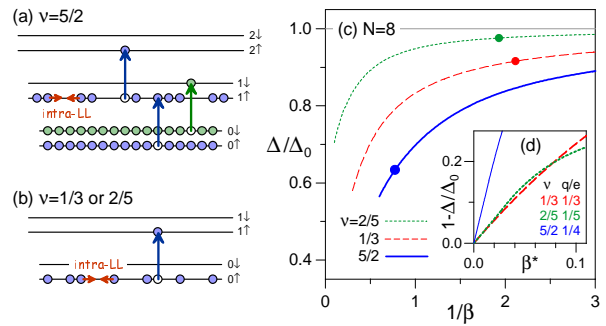}
\caption{(color online)
   (a,b) Comparison of inter-LL excitations with up to one 
   unit of cyclotron energy in the $\nu={5\over2}$ state 
   and in Laughlin/Jain liquids at $\nu<1$.
   (c) Reduction of the excitation gap of different quantum 
   Hall states due to LL mixing, plotted as a function 
   of the cyclotron-to-Coulomb energy ratio $\beta^{-1}$.
   $\Delta$ and $\Delta_0$ are gaps calculated for $N=8$ 
   and $2l=16$, 21, and 17 (for $\nu={2\over5}$, ${1\over3}$,
   and ${5\over2}$).
   The full dots correspond to the electron concentration 
   $\varrho=2.3\times10^{11}$~cm$^{-2}$.
   (d) Gap reduction $1-\Delta/\Delta_0$ as a function 
   of parameter $\beta^*$ in which Coulomb energy 
   $q^2\lambda_q^{-1}$ involves different fractional 
   charge quanta $q$ (indicated) appropriate for 
   different liquids.}
\label{fig2}
\end{figure}

To include LL mixing, we expand the Hilbert space by adding the 
$K=1$ states, i.e., allow excitation of up to one electron to a 
higher LL.
This can be regarded as a kind of first-order perturbation scheme 
in $\beta$.
However, note that such coupling of the entire $K=0$ and $K=1$ 
subspaces is more than a first-order perturbation on the specific 
$K=0$ eigenstates.
The basis is shown schematically in Fig.~\ref{fig2}(a) and (b).
At $\nu={5\over2}$, the underlying, filled lowest LL gives rise to 
more types of excitations than at $\nu<1$.
Generally, the inter-LL excitations can be decomposed into addition
of an electron or a hole to a specific LL in the presence of a 
correlated state of the initial $N$ electrons.
There are two distinct cases, depending on the target LL:
(i) 
Addition to (or removal from) the LL occupied by the incompressible 
liquid.
This causes creation of several fractionally charged QPs.
At $\nu={1\over3}$ or ${2\over5}$ they are well known Laughlin or 
Jain QPs, with a simple and intuitive picture in the CF model.
However, the nature and dynamics of the QPs at $\nu={5\over2}$ are not
nearly as well understood.
(ii) 
Addition to (or removal from) a different LL.
This makes the added/removed electron distinguishable from the 
correlated electrons.
This problem resembles coupling of a Laughlin liquid to a (positive) 
valence hole\cite{Chen93} or a (negative) trion.\cite{QX}
However, coupling of the $\nu={5\over2}$ state to a foreign charge 
is far less understood.

Since all three types of $K=1$ excitations must be included in 
the calculation on the same footing, even this limited account 
of LL mixing boosts the space dimension from $\sim10^3$ to 
$\sim4\times10^5$ for $N=8$ at $2l=17$.
This precludes similar calculations for larger systems or further 
inclusion of the $K>1$ excitations.
On one hand, this makes the present results somewhat susceptible
to finite-size errors (although Fig.~\ref{fig1}(b) may suggest
that $N=8$, i.e., four pairs, is already a representative system)
and accurate only up to the first-order perturbation in $\beta$.
On the other hand, a much larger number of $K=1$ excitations at 
$\nu>2$ than at $\nu<1$ suggests that the effects of LL mixing should 
be more important at $\nu={5\over2}$ than in Laughlin or Jain liquids
of the lowest LL.

\section{Results and discussion}

We focus on two features of the $\nu={5\over2}$ state: the excitation 
gap and the overlap with the MR wavefunction.
The results of calculations of the gap $\Delta$ for $N=8$ and 
$K\le1$ in a 2D layer of zero width are shown in Fig.~\ref{fig2}(c).
As argued above and anticipated from experiments, the gap reduction
$\Delta/\Delta_0$ (where $\Delta_0$ is the result for $K=0$) is 
noticeably greater at $\nu={5\over2}$ than at both $\nu={1\over3}$ 
or ${2\over5}$.
On the other hand, the fact that the gap at $\nu={1\over3}$ is 
reduced more than at $\nu={2\over5}$ can be related to the smaller 
QP charge in the latter case ($q=e/5$ versus $e/3$).
Since the low-energy response of a liquid involves formation 
and interaction of the QPs, the harsh $\beta\ll1$ criterion 
for the accuracy of the isolated-LL approximation may be relaxed 
to $\beta^*\ll1$.
Here, 
$\beta^*=(q^2\lambda_q^{-1})/(\hbar\omega_c)=(q/e)^{5/2}\beta$ 
involves the Coulomb energy scale of the QPs.
Indeed, when the gaps in the weak-perturbation regime are plotted 
as a function of $\beta^*$ as in Fig.~\ref{fig2}(d), the data 
for $\nu={1\over3}$ and ${2\over5}$ falls close to the same line,
$1-\Delta/\Delta_0\approx3\beta^*$.
Taking $q=e/4$ for the $\nu={5\over2}$ state results in a much
(about $3\times$) steeper curve.
This indicates that the response of the $\nu={5\over2}$ state to 
the perturbation associated with the LL mixing at a finite $\beta$
is (due to a richer inter-LL excitation spectrum) relatively stronger
than the response of Laughlin or Jain states in the lowest LL.

Furthermore, if the experiments on all three electron liquids were 
to be carried out at similar concentrations (corresponding to a maximum 
mobility), the difference between them will be additionally magnified 
by a difference in $\beta$ corresponding to different $\nu$.
For example, for $\varrho=2.3\times10^{11}$~cm$^{-2}$ we obtained
gap reduction of 8.5\%, 2.5\%, and 35\% at $\nu={1\over3}$, 
${2\over5}$, and ${5\over2}$, respectively.

In the above discussion we have established the following.
(i) Realistic estimates of the excitation gap at $\nu={5\over2}$
must include the LL mixing, whose effect at this filling is much 
stronger than for Laughlin or Jain states in the lowest LL.
(ii) The gap reduction caused by LL mixing is already significant 
in the $K\le1$ approximation.
(iii) It is plausible that the full account of the LL mixing might 
reconcile experimental results in the limit of vanishing disorder 
with the numerics. 
Unfortunately, calculations for $K>1$ and $N\ge8$ are beyond 
our present capabilities.

\begin{figure}
\includegraphics[width=3.4in]{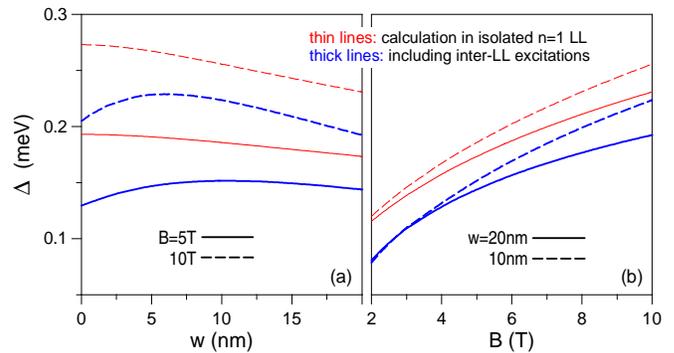}
\caption{(color online)
   Dependence of the excitation gaps $\Delta$ calculated at 
   $\nu={5\over2}$ (for $N=8$) with and without LL mixing, on
   the 2D layer width $w$ (a) and the magnetic field $B$ (b).}
\label{fig3}
\end{figure}

The dependence of gap $\Delta$ on the magnetic field $B$ and 
finite layer width $w$ are displayed in Fig.~\ref{fig3}.
Finite width was introduced to the model by the calculation
of two-body Coulomb matrix elements using 3D wavefunctions
$\chi(z)\psi_{nm}(\theta,\phi)$ with $\chi(z)\propto\cos(\pi 
z/w)$ for the normal direction.
Remarkably, $\Delta(w)$ is nonmonotonic, with a maximum between 
$w=5$ and 10~nm, depending on $B$.
While the reduced gaps are still about twice larger than the 
experimental values, the inclusion of even only $K=1$ excitations 
clearly improves the model.
This suggests LL mixing as the main reason for the earlier 
$\Delta$-discrepancy.

Let us now turn to the question of equivalence of the $\nu=
{5\over2}$ state realized in experiment and the model MR 
wavefunction.
There is a subtle difference between the half-filled state and 
odd-denominator liquids like $\nu={1\over3}$ or ${2\over5}$.
In the latter states, it is not merely incompressibility but also
the form of correlations and many-body wavefunction that are robust 
against the variation of material, $w$, or $B$ --- as long as the 
interaction pseudopotential is sufficiently strong (superharmonic) 
at short range, weak compared to the cyclotron energy, and strong 
compared to disorder.
In contrast, the half-filled state remains incompressible for a
wide class of electron--electron pseudopotentials, but the exact
form of the wavefunction strongly depends on their details.
This limits information that can be inferred about the nature 
of the state from the observation of its incompressibility.

The underlying reason is the competition of at least two distinct 
wavefunctions sharing the same symmetry: the MR state, which can be 
defined as an exact ground state of a short-range three-body 
repulsion,\cite{Greiter91} and a clustered state characterized
\cite{clusters} by the maximum avoidance of the next to the lowest 
value of the relative pair angular momentum, $\mathcal{R}=3$.
The MR state is anticipated for the pair repulsion which is nearly 
harmonic at short range (i.e., with the pseudopotential decreasing 
linearly through $\mathcal{R}=1$, 3, and 5), such as in the LL$_1$.
However, the overlaps of the actual Coulomb eigenstates obtained
from finite-size numerics (in the $K=0$ approximation) with the 
MR state are sensitive to the interaction parameters\cite{Morf98}
and surface curvature,\cite{3body} raising the question of whether 
the $\nu={5\over2}$ FQH state and the MR model state are indeed 
(qualitatively) equivalent.

\begin{figure}
\includegraphics[width=3.4in]{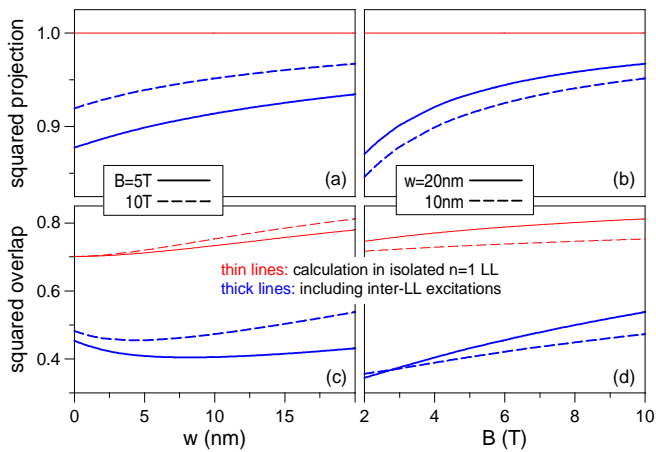}
\caption{(color online)
   Dependence of the squared projection onto the LL$_1$ 
   (top) and the squared overlap with the Moore--Read
   wavefunction (bottom) of the $\nu={5\over2}$ ground 
   states calculated for $N=8$ with and without LL mixing, 
   on the 2D layer width $w$ (left) and the magnetic 
   field $B$ (right).}
\label{fig4}
\end{figure}

How does LL mixing affect this problem?
In the top frames of Fig.~\ref{fig4} we plot the squared projection 
$|\mathcal{P}_{K=0}\Psi|^2$ onto the LL$_1$ (i.e., onto the $K=0$ 
subspace) of the same ground states $\Psi$ whose gaps are  shown 
in Fig.~\ref{fig4}.
It depends on $w$ and $B$, but it is always significantly higher 
than $\Delta/\Delta_0$ or the (not shown) squared overlap with the 
$K=0$ ground state, in consequence of the coupling between intra- 
and inter-LL excitations in the $K\le1$ space.
In the bottom frames we plot the squared overlap 
$\xi^2=\left|\left<{\rm MR}|\Psi\right>\right|^2$ with the MR state 
(more precisely, with its particle--hole conjugate at $2l=2N+1$).
The small values for the $K=0$ calculation (here, 0.75 to 0.80) may 
be to some extent an artifact of spherical geometry.\cite{3body}
However, a significant drop caused by the LL mixing (e.g., from
0.78 to 0.43 for $w=20$~nm and $B=5$~T) suggests that the MR 
wavefunction may not be a very realistic description of the 
$\nu={5\over2}$ state in this range of parameters.
This ambiguity and the difficulty with more realistic calculations 
make further experiments\cite{Stern06} irreplaceable.

\section{Conclusion}

We studied the effect of LL mixing on the $\nu={5\over2}$ FQH state.
This effect is more pronounced than in the incompressible liquids of 
the lowest LL such as $\nu={1\over3}$ or ${2\over5}$ (even at the same 
magnetic field) due to the richer inter-LL excitation spectrum.
The resulting reduction of the excitation gap explains the troubling 
disparity between previous numerics (without LL mixing) and the 
experiments.
This prediction agrees qualitatively with Ref.~\onlinecite{Morf03} 
which included LL mixing in a different way.
Finally, the LL mixing significantly lowers overlaps with the MR 
wavefunction.
This amplifies the need for the recently proposed\cite{Stern06} 
experiments, designed to probe directly the nonabelian statistics 
of the QPs at $\nu={5\over2}$.


The authors thank Wei Pan for helpful discussions and acknowledge
support from grants DE-FG 02-97ER45657 of US DOE and N20210431/0771 
of the Polish MNiSW.

\end{document}